\documentclass[preprint,pra,aps,floatfix,showpacs]{revtex4}
\usepackage{epsfig,graphicx,tabularx}

\begin{document}
\title{ Winding number dependence  of Bose-Einstein condensates in a ring-shaped lattice}
\author{D. M. Jezek and H. M. Cataldo}
\affiliation{IFIBA-CONICET
\\ and \\
Departamento de F\'{\i}sica, FCEN-UBA
Pabell\'on 1, Ciudad Universitaria, 1428 Buenos Aires, Argentina}
\date{\today}
\begin{abstract}
We study the winding number dependence of the stationary states of a
Bose-Einstein condensate in a ring-shaped lattice.
The system is obtained by confining  atoms in a toroidal trap 
with equally spaced radial  barriers. 
We  calculate the energy and  angular momentum
as functions of the winding number and the barrier height for two quite distinct 
particle numbers.
In both cases we observe two clearly differentiated regimes. 
For low barriers, metastable vortex states are
obtained up to a maximum winding number which depends on the  particle number and 
 barrier height.
In this regime, the angular momentum and energy show, respectively,
 almost linear and quadratic dependences on the winding number.
For large barrier heights, on the other hand,  stationary states are obtained 
up to a maximum winding number which
depends only on the number of lattice sites,
whereas  energy and angular momentum  are shown to be sinusoidal functions
of the winding number.

\end{abstract}

\maketitle
\section{Introduction}

Ultra-cold atoms in optical lattices nowadays represent
a perfect laboratory system
for the quantitative study of various condensed-matter
models. In addition to
offering an extremely high degree
of experimental control and reproducibility, such systems have shown to provide
 a very useful theoretical testing ground
for the quantum modeling of strongly
correlated many-body systems. Perhaps the
most simple and fecund of such applications is given by
the dynamics of double-well Bose-Einstein
condensates \cite{gati}, which has been extensively studied in recent years
\cite{twowell}.
In particular, for high enough barriers, 
a simple two-mode model Hamiltonian
describing Josephson-oscillations
and quantum self-trapping phenomena,
has been analyzed by different authors \cite{Smerzi1997,Raghavan1999,zap98,zap03},
and later improved
 to better introduce interparticle interaction effects \cite{Ananikian2006}.
On the other hand, the associated dynamics  has been experimentally
observed a few years ago \cite{Albiez2005}.

In the last years, this research work has been extended to 
 multiple well condensates motivated by the possibility of generating and
manipulating   vortex states.
In a recent experiment, Scherer {\it et al.} \cite{scher07} have reported observations
of vortex formation by merging and interfering multiple Bose-Einstein condensates
in a confining potential. In their experiment, a single harmonic potential well was
partitioned into three sections by barriers forming three independent Bose-Einstein
condensates, which had no coherence between the respective phases. 
By removing the barriers, the condensates merged together and it has been shown 
that, depending on the unknown initial
relative phases, the final state could acquire  vorticity. 
The authors have observed single vortex states even after five seconds
following  the barrier ramp-down, indicating relatively long single vortex lifetimes.
This experiment stimulated  theoretical investigation  as regards the
maximum winding number a condensate in a ring-shaped lattice  could sustain,
depending on the trap properties, particularly, on the barrier height between wells. 
For example, in the case of large barrier heights, the problem of vortex trapping 
in cyclically coupled Bose-Josephson
junctions has been recently addressed by Ghosh and Sols \cite{ghosh08}.
They coupled initially independent Bose-Einstein condensates 
through Josephson links and allowed the system to reach a stable circulation
by adding a dissipative term in semiclassical equations of motion.
In addition, the authors
analyzed the probability 
of trapping a vortex with a given 
winding number, finding that a necessary condition to obtain a metastable
nonzero circulation consists of generating a winding number smaller than a quarter 
of the total number of
linked condensates. A different approach was utilized by
P\'erez-Garc\'{\i}a {\it et al.} \cite{gar07}, which by means of group-theoretical
methods,
studied charge inversion  and vortex erasing in condensates
confined by  traps  with discrete symmetries \cite{fer05}. 
They have predicted
the existence of
stationary states up to a   winding number which depends on the order of the 
symmetry. 
On the other hand, the case of a continuous symmetry, represented by 
nonrotating ring shaped traps, has attracted renewed theoretical attention
 in connection with the existence of 
vortex states, due to recent experimental observations of persistent currents
in multiple connected 
condensates \cite{ryu07,wei08}.
In particular,  by computing  the energy landscape,
  Capuzzi {\it et al.} \cite{cap09} have shown   
 that multiply
quantized vortices and multivortex configurations can be locally energetically stable
 in a toroidal trap.

The aim of this work is to study vortex states in a ring-shaped optical
 lattice with variable barrier heights. This will allow us to explore
the transition between the limit of low barrier heights, where large persistent currents are expected,
and the high-barrier limit, where  Josephson-type links should rule the vortex configurations. 
For such purpose we will consider a condensate confined 
by superposition of
 a toroidal trap and
equally spaced radial barriers forming a ring of $N_c$ linked   condensates.
In particular, we will investigate the energy and angular momentum dependence on the winding number
for a wide range of barrier heights.
Such a  system could be experimentally investigated, as it corresponds to
a relatively simple combination of  previously implemented experimental 
settings \cite{scher07,wei08}. Actually, new techniques have been recently developed by Henderson 
{\it et al.} \cite{hend} to create and manipulate condensates in a variety of geometries, including
the present type of ring lattices.

Our work is organized as follows, in Sec.~\ref{Theory} we describe the
theoretical framework utilized to generate stationary states
with different winding numbers. Sec.~\ref{Models} is devoted to
an analytical study of the energy and the angular momentum for limiting values
 of the barrier height.
By numerically solving the Gross-Pitaevskii equation,
we calculate in Sec.~\ref{States}  the energy and angular momentum and analyze their
dependence on the winding number and the barrier height.
Finally, a summary and concluding remarks are
offered in Sec.~\ref{Conclusions}.

\section{Theoretical Framework}\label{Theory}
 
 \subsection{The Trap}

\begin{figure}
\includegraphics{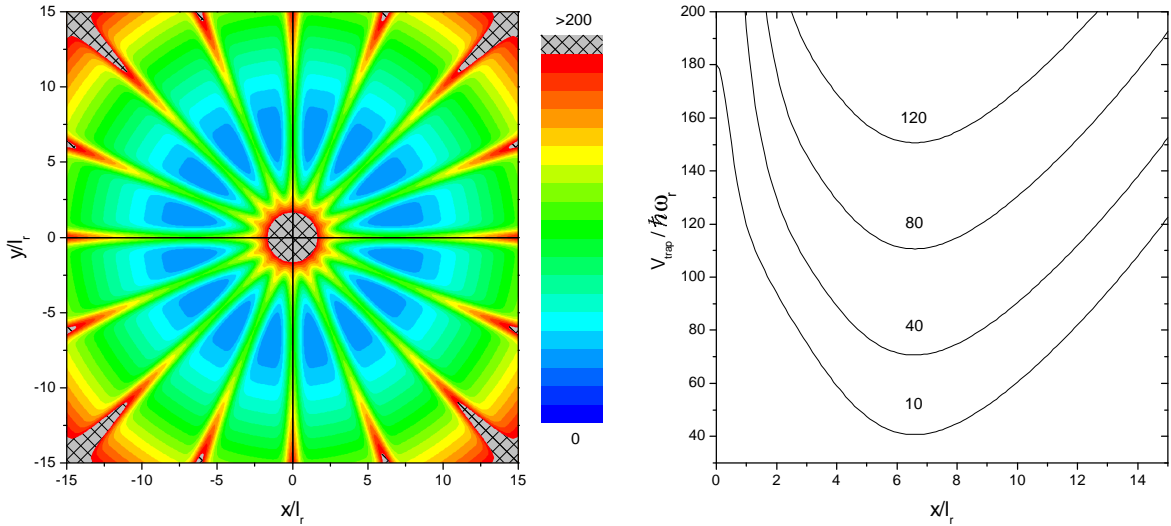}
\caption{(Color online) 
Left panel: equipotential curves for the external trap  $ V_{\text{trap}}(x,y)/\hbar \omega_r$
with $V_b/\hbar \omega_r=80 $. Right panel:
effective potential barrier $ V_{\text{trap}}(x,0)$ along the positive $x$ axis 
for  several values of the  parameter
 $V_b/\hbar \omega_r $.}
\label{potenr}
\end{figure}

We consider  a Bose-Einstein condensate  of Rubidium atoms confined by 
an external trap $ V_{\text{trap}}$, consisting of a superposition of
a toroidal term $V_{\text{toro}}$  and a lattice potential  $V_{\text{L}}(x,y)$
 formed by radial barriers.
Similarly to the trap 
utilized in  recent experiments \cite{ryu07,wei08}, the toroidal trapping  potential
 in cylindrical coordinates reads,
\begin{equation}
V_{\text{toro}}(r,z ) = \frac{ m }{2 } \left[\omega_{r}^2  r^2 
+ \omega_{z}^2  z^2\right] +  
V_0 \, \exp ( -2 \, r^2/\;\lambda_0^2)
\end{equation}
where $\omega_{r}$  and $\omega_{z}$ denote the radial and axial 
frequencies, respectively. 
We have set $\omega_z >>
\omega_r$ to suppress excitation in the $z$ direction.
In particular, we have chosen
 $ \omega_r / (2 \pi) =  7.8 $ Hz  and $ \omega_z / (2
\pi) = 173 $ Hz, while for the laser beam we have set 
$ V_0  =  100  \, \hbar \omega_r$  and $ \lambda_0 = 6 \, l_r $, with 
$ l_r = \sqrt{\hbar /( m \omega_r)}$. 
On the other hand, the lattice  potential formed by $N_c$
Gaussian  barriers located at equally spaced angular positions $ \theta_i = 2 \pi i /N_c $
($0\leq i\leq N_c -1$) may be written,
\begin{equation}
V_{\text{L}}(x,y) =
V_b \,\, \sum_{i=0}^{N_c/2-1} \exp \left\{ - \frac{ [ \cos(\theta_i) \, y  - \sin(\theta_i) \,x]^2}
{ \lambda_b^2}\right\}
\end{equation}
for $ N_c$ even. Although, for simplicity the numerical simulations of this work
have been restricted to the case of an even number of lattice sites,
most of our findings can be easily shown to be valid irrespective of the parity of $ N_c$.
We have fixed the trap parameters at 
$ \lambda_b=0.5 l_r $ and $ N_c = 16 $. 
The combination of these potentials gives rise to effective barriers between neighboring
sites, as represented in Fig.~\ref{potenr}. 

The Thomas-Fermi (TF) approximation, will allow us to deduce
some general features of the condensate from the characteristics of the trapping potential.
To this aim, we will take into account
that the TF-density vanishes at points fulfilling
$ \mu < V_{\text{trap}}(x,y)$,
$ \mu $ being the chemical potential. On the one hand, around the trap center we 
can approximate
$ V_{\text{trap}} \simeq V_0 + (N_c/2)  V_{b} $  and,
as we restrict our calculations to  chemical potentials
 smaller than $  V_0$,  the condensate center will always exhibit a hole.  
 On the other hand, between neighboring sites
we have that the potential minimum shown in the right panel of  Fig.~\ref{potenr}, 
which corresponds to the saddle point in two dimensions,
turns out to be located around  $ x_{\text{min}}=6.6\, l_r $, and the  barrier height 
can be approximated as 
 $ V_{\text{min}} \simeq V_{\text{trap}}(6.6,0) = 30.7 \, \hbar \omega_r +V_{b}$.
Notice that we have retained in this estimate only the $i=0$ term of the sum  in $V_{\text{L}}(x,y)$,
as the exponent  of such a term  vanishes, while the contributions
of the remaining Gaussian terms are negligible.
Thus, when the chemical potential becomes smaller than this 
potential minimum, the TF-density between neighboring sites  vanishes. 

\subsection{Stationary States}

\begin{figure}
\includegraphics{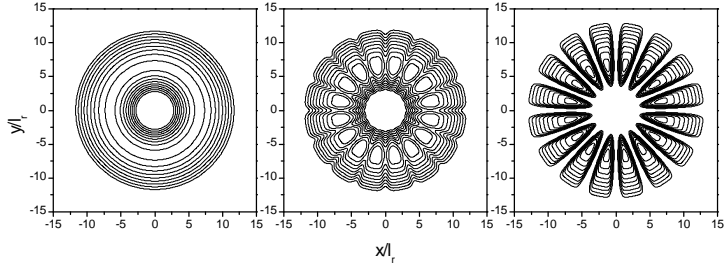}
\caption{ Density isocontours
for several values of  the barrier  parameter. From left to right:
 $ V_b/\hbar \omega_r = 0, 10, $ and $ 80 $. The number of particles corresponds to
$ N= 10^5 $.
}
\label{fun}
\end{figure}

Since the trapping potential is much stronger along the $z$ axis,
 the motion of the particles in the $z$ direction
remains frozen in the ground state, and the energy per particle in the
condensate is then given by
the two-dimensional (2D) energy
 functional \cite{castin}
\begin{equation}
  E [\psi]  =\int \left( \frac{ \hbar^2 }{2 m}  |\nabla \psi |^2 +
V_{\rm{trap}} \,|\psi|^2 + \frac{1}{2} N g \, |\psi|^4  \right) dx\, dy ,
\label{ed}
\end{equation}
where $ \psi=e^{i\phi}|\psi|$ denotes the 2D order parameter, $N$ the number of particles, and $m$  the atom
mass. The effective 2D coupling constant $g=g_{3D}\sqrt{m\omega_z/2\pi\hbar} $
  is written in terms of the 3D coupling constant between the atoms 
$g_{3D}=4\pi a\hbar^2/m$, where $a= 98.98\, a_0 $ denotes the  
$s$-wave scattering length of $^{87}$Rb, $a_0 $ being the Bohr radius.
In order to obtain stationary states, one has to perform
variations of $ E $ with respect to $ \psi $ keeping the number of
particles   fixed,  which yields the Gross-Pitaevskii (GP) equation \cite{gros61}
\begin{equation}
 \left (-\frac{ \hbar^2 \nabla^2}{2 m}+ V_{\rm{trap}} 
+N g \, |\psi|^2 
 \right ) \, \psi = \mu \psi \; .
\label{gp}
\end{equation}
The ground state wavefunction  is numerically obtained without imposing any constraint.
Typical configurations are shown in Fig. \ref{fun}, where we depict the equidensity 
curves for  different  barrier heights.
  The left panel corresponds to
the barrier-free system, while the middle and right panels correspond to
barrier parameters fixed at $ V_b/\hbar \omega_r= 10 $ and $V_b/\hbar \omega_r=80$, 
respectively.
 We note that the chemical potential verifies
$ \mu \simeq 77\, \hbar \omega_r > V_{min}  $ for the lower barrier,
 while the higher barrier   verifies the opposite inequality 
$ \mu \simeq 90\, \hbar \omega_r  < V_{min}$. 
Thus, as we have explained in the previous section, the TF-density between 
neighboring sites  should vanish in the latter configuration,
while remaining finite in the former one. As regards the exact density, it should exhibit  a similar behavior,
except for a smoother decay.

On the other hand,  
further constraints must be imposed in order to obtain stationary states with nonvanishing winding numbers,
and to this aim we have employed  a method which was  described in detail in a previous work \cite{jez08}.
Summarizing, we begin by using the ground state wavefunction $\psi_0$
to construct an initial
order parameter with vorticity \cite{cap09}:

\begin{equation}
\psi({\bf r})=\psi_0({\bf r}) \,( \frac{x+iy}{\sqrt{x^2+y^2}})^n \,,
\label{wf-ansatz}
\end{equation}
where  $n$ denotes the initial winding number or topological charge. 
From this ansatze, it is straightforward to realize that $\psi$ and
$\psi_0$  possess the same density profile, but $\psi$ presents an imprinted
velocity field. Such a field is irrotational everywhere, except at the origin,
with a circulation along any closed loop around this point, which must be proportional to the winding number $n$.
Next, we apply a minimizing procedure  in order to  obtain a stationary
state.
During the minimization  such a velocity field undergoes drastic changes when
large barrier heights are considered.
Notice that the presence of radial barriers destroys the axial symmetry,  and thus
the  angular momentum 
does no longer commute with the Hamiltonian. As a consequence, there is also no longer
a linear relationship between the value of the angular momentum and
 the imprinted winding number. In addition, the final winding number $n'$ 
after the minimization process may differ
from the initial one $n$, as will be shown in Sec.~\ref{States}, where
we will also see that $ |n'| $ is bounded by a maximum value, which we shall call $ \nu $.

\section{ Analytical expressions }\label{Models}

In this Section, we shall  analyze the  behavior of the angular momentum and the energy as
 functions of the winding number in two limiting cases, namely in the
absence of   barriers   and for
 large barrier heights. In both cases we will derive analytical 
expressions for the mean values
of such observables in terms of  ground state magnitudes. 
Hereafter, when referring to energy and angular momentum, it must be understood that
they denote the mean value of the corresponding observables per particle.

\subsection{ Barrier free system }\label{sinbarrera}

Taking into account that the ground state density exhibits a hole at the condensate center,
for vanishing barriers and an arbitrary winding number $n$,
 the order parameter 
may be approximated as:

\begin{equation}
\psi_n( r, \theta ) = \psi_0( r) \, e^{i n \theta } \,,
\label{wfi0}
\end{equation}
where we have assumed that
the presence of a non-homogeneous phase in the wavefunction with its 
concomitant velocity field, does not
affect appreciably the shape of the condensate density, as compared  to that of  the ground state.
This occurs because the centrifugal force should have 
little effect outside the central hole,  where the density is non-negligible.

Replacing the wavefunction (\ref{wfi0}) in the first term of the integrand of 
(\ref{ed}) we have:

\begin{equation}
\frac{ \hbar^2 }{2 m}  |\nabla \psi_n |^2 = \frac{ \hbar^2 }{2 m}  (\nabla \psi_0 )^2 
+ \frac{ \hbar^2 }{2 m}  ( \psi_0 )^2  \frac{ n^2 }{ r^2} 
\label{cine}
\end{equation}
and one can easily obtain the  mean value of the energy as a quadratic function
of the winding number $n$,

\begin{equation}
E(n)= K n^2 + E_0
\label{ev0}
\end{equation}
where $ E_0$ denotes the ground state energy and $ K $ is given by:

\begin{equation}
K = \frac{\hbar^2}{2m}  \, 2 \pi \int \frac{ 1}{ r^2 } [\psi_0( r)]^2 r dr.
\label{k}
\end{equation}

On the other hand, since the angular momentum  commutes with the Hamiltonian
and  the wavefunctions (\ref{wfi0}) are eigenfunctions of $ \hat{L}_z $, 
we have the linear relationship

\begin{equation}
L_z=\hbar n  \,.
\label{lz00}
\end{equation}

\subsection{ High barriers }\label{barrerasaltas}

\begin{figure}
\includegraphics{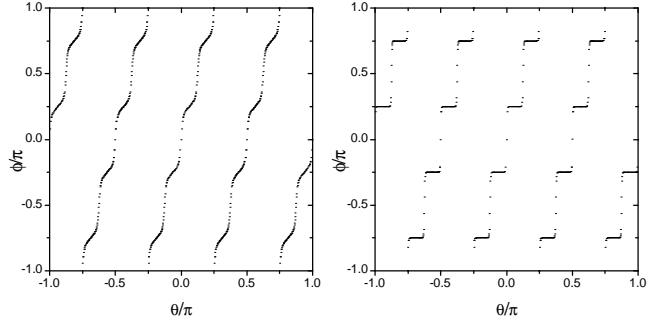}
\caption{  Phase of the order parameter $\phi$ as a function of the angular coordinate for $ r_0=6\, l_r $,
a winding number $n=4$, and two values of the barrier height $V_b/\hbar \omega_r=60$ (left panel)
  and $V_b/\hbar \omega_r=120$ (right panel).
The number of particles is $N=10^5$.}
\label{phase5}
\end{figure}

According to  Refs. \cite{fer05} and \cite{gar07} the most general solution of the GP equation is
a Bloch state of the form:

\begin{equation}
\psi_n( r, \theta ) = e^{i n \theta } \, f_n ( r, \theta ) \,,
\label{wfbloch}
\end{equation}
where the function $ f_n ( r, \theta )$ is invariant under a $ 2 \pi /  N_c $ rotation in 
$\theta $.
This type of solution is valid irrespective of the barrier height. The above periodicity
is obvious in the modulus, since  $ |f_n ( r, \theta )|^2= |\psi_n ( r, \theta )|^2 $
corresponds to the periodic particle density as shown in Fig.~\ref{fun}.
However, the phase of $ \psi_n ( r, \theta )$ does not necessarily present this  symmetry.
 As an example, in Fig.~\ref{phase5}  
we depict the phase of $\psi_4( r_0, \theta )$ as a function of the angular coordinate,
which clearly does not exhibit such a symmetry.
Whereas, it is easy to verify that the phase of
 $ f_4 ( r_0, \theta )= e^{-i 4 \theta}\psi_4( r_0, \theta )$ is indeed a $ 2 \pi /  N_c $ periodic function.

In this section we will concentrate ourselves on  large  barrier heights.
In such a limit, the phase turns out to be almost uniform in each site, as it may be seen 
from the right panel of Fig.~\ref{phase5}. This is a consequence of the fact that
for high barriers, the condensate portions become more isolated and 
thus the
particle current diminishes. This, in turn, implies that the velocity field $\vec{v}$ and hence the gradient
of the phase
$ (m/ \hbar) \vec{v}$ practically vanishes within each portion.
Then, for large enough barriers, such that the chemical potential verifies $ \mu < V_{min} $,
the system  splits into  a ring of  weakly linked condensates, and
the order parameter
can be approximated as follows,

\begin{equation}
\psi_n({ r, \theta })= \sum_{k=0}^{N_c-1} \Psi_k({ r, \theta}) \, e^{i  2 \pi n k / N_c  } \,,
\label{wfii}
\end{equation}
where $ \Psi_k({ r, \theta})$ is a   real function describing a localized state in the $k$-site,
 i.e., having only a very little overlap
with neighboring sites.
This approximation is analogous to the tight-binding approximation \cite{ash}, and
the  matrix element $\psi_n^* \hat{\cal{O}}  \psi_n $ of a given  observable 
 $ \hat{\cal{O}} $ can be approximated as

\begin{equation}
  \sum_{k=0}^{N_c-1} \Psi_k \,
 \hat{\cal{O}}   \, \Psi_k   \,
+ e^{i \frac{ 2 \pi n}{ N_c} } \sum_{k=0}^{N_c-1} \Psi_k \,
  \hat{\cal{O}} \, \Psi_{k+1}  \,
+  e^{-i \frac{ 2 \pi n}{ N_c}  }\sum_{k=0}^{N_c-1} \Psi_k \,
  \hat{\cal{O}}  \, \Psi_{k-1}  \,,
\label{htermsgen}
\end{equation}
where we have assumed that only the matrix elements between neighboring sites are
non-negligible. In our notation, the site denoted by $k=N_c$ ($k=-1$) is 
identified with that of $k=0$
($k=N_c-1$). 
Then, as the summation is performed over all $k$ values, we can rewrite 
the expression (\ref{htermsgen}) as follows:

\begin{equation}
  \sum_{k=0}^{N_c-1} \Psi_k \,
 \hat{\cal{O}}   \, \Psi_k   \,
+ e^{i \frac{ 2 \pi n}{ N_c} } \sum_{k=0}^{N_c-1} \Psi_k \,
  \hat{\cal{O}} \, \Psi_{k+1}  \,
+  e^{-i \frac{ 2 \pi n}{ N_c}  }\sum_{k=0}^{N_c-1} \Psi_{k+1} \,
  \hat{\cal{O}}  \, \Psi_{k}  \,.
\label{optermsgen1}
\end{equation}

Let us now assume that the observable $ \hat{\cal{O}} $ is the  angular momentum operator:

\begin{equation}
  \hat{\cal{O}}    \, = \hat{L}_z= \frac{\hbar}{i}  \, \frac{\partial}{\partial \theta}.
\label{mom1}
\end{equation}
 Using that $\hat{L}_z$ is hermitian and $ \Psi_k $ are real functions, the following 
equalities hold:

\begin{equation}
 \Psi_k \, \hat{L}_z  \, \Psi_{k}  \, =  0
\label{mom12}
\end{equation}

\begin{equation}
 \Psi_k \, \hat{L}_z  \, \Psi_{k+1}  \, = -
 \Psi_{k+1} \, \hat{L}_z  \, \Psi_{k}  \,
\label{mom2}
\end{equation}
Then replacing (\ref{mom12}) and (\ref{mom2})  in (\ref{optermsgen1})
we obtain 
\begin{equation}
\psi_n^* \hat{L}_z \psi_n = 
\frac{\hbar}{i} (e^{i \frac{ 2 \pi n}{ N_c}}  -
  e^{-i \frac{ 2 \pi n}{ N_c}} ) \sum_{k=0}^{N_c-1} \Psi_k
  \frac{ \partial \Psi_{k+1}}{ \partial \theta}  \,
\label{aterms3}
\end{equation}
and thus the mean value of the angular momentum reads:

\begin{equation}
\int \, \psi_n^* \hat{L}_z \psi_n \, r \, dr \, d\theta = 
 2 \, \hbar \, \sin(2 \pi n/ N_c)  \sum_{k=0}^{N_c-1} \, \int \, \Psi_k
  \frac{ \partial \Psi_{k+1}}{ \partial \theta} \, r \, dr \, d\theta \,.
\label{angf}
\end{equation}
Note in the above expression that the winding number appears only in the argument of the sine,
 which determines that the
angular momentum must exhibit a sinusoidal behavior as a function of $n$.

In computing the mean value of the energy, it is convenient to separately treat the
interacting and noninteracting terms. In fact, the noninteracting part, which corresponds
to the observable $ \hat{\cal{O}}_s = -\frac{ \hbar^2 \nabla^2}{2 m}+ V_{\rm{trap}}$,
may be treated analogously to the angular momentum, 
yielding

\begin{equation}
 \Psi_k \, \hat{\cal{O}}_s \, \Psi_{k+1}  \, = 
 \Psi_{k+1} \,  \hat{\cal{O}}_s \, \Psi_{k}  \,,
\label{ops}
\end{equation}
which replaced in  (\ref{optermsgen1})  leads to

\begin{eqnarray}
& &\int\int\psi^*_n \, \hat{\cal{O}}_s  \, \psi_n  \, \,
r\,dr d\theta
 = \sum_{k=0}^{N_c-1}\int\int \Psi_k \,
\hat{\cal{O}}_s   \, \Psi_k \, \,
r\,drd\theta\nonumber \\
& + & 2 \,\cos(2\pi n/N_c)\sum_{k=0}^{N_c-1}\int\int \Psi_k \,
\hat{\cal{O}}_s \  \, \Psi_{k+1} \, \,
r\,drd\theta.
\label{20}
\end{eqnarray}
On the other hand, the interacting term corresponds to the nonlinear operator
$ \hat{\cal{O}}_{nl} = \frac{g}{2}N|\psi_n|^2$, proportional to the particle density,
for which the approximation (\ref{optermsgen1}) is no longer valid. 
In this case, the mean value  must be 
computed to first order in the product of adjacent wavefunctions $\Psi_k\Psi_{k+1}$,
yielding
\begin{eqnarray}
& &\frac{g}{2}N\int\int|\psi_n |^4\,
r\,dr d\theta
 = \sum_{k=0}^{N_c-1}\int\int \frac{g}{2}N \Psi_k^4\,
r\,drd\theta\nonumber \\
& + & 2 \,gN\cos(2\pi n/N_c)\sum_{k=0}^{N_c-1}\int\int (\Psi_k^3\Psi_{k+1}+
\Psi_k\Psi_{k+1}^3) \,
r\,drd\theta.
\label{21}
\end{eqnarray}
Finally, from (\ref{20}) and (\ref{21}) we may conclude that  the mean value of the
energy will present a  dependence of the type $ \cos(2\pi n/N_c)$ on
the winding number.

\section{Energy and angular momentum: Numerical Results }\label{States}

\subsection{ Low barriers}\label{lowb}

A vanishing barrier ($ V_b =  0 $) yields a
ground state  chemical potential 
$ \mu = 33.19 \, \hbar \omega_r $ ($\mu =73.31 \, \hbar \omega_r $)
 for  $ N=10^3 $  ($ N=10^5 $) particles. Such values turn out to be
  smaller than $ V_0  $, and thus, as discussed in Section II,
 the corresponding condensates
 exhibit a hole at their centers \cite{cat09}. It has been shown, both experimentally
\cite{ryu07} and theoretically \cite{cap09}, that this  type of configuration 
forming a torus
 can host locally stable multiply quantized vortices.
We have found such stable 
vortices up to a maximum winding number of $\nu =7$ ( $ \nu =9$ ) for $ N=10^3 $ ($ N=10^5 $),
reflecting the fact that $\nu$  depends on the number of particles. Actually, we will next see that 
$\nu$ also depends on the barrier height.

\begin{figure}
\includegraphics{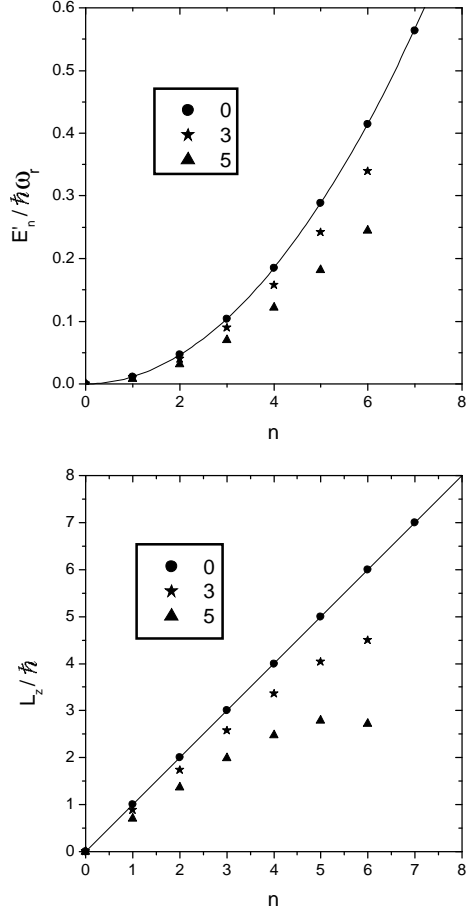}
\caption{   Excitation energy per particle $E'_n$ (top panel), 
and angular momentum per particle $L_z=\langle \hat{L}_z\rangle$ (bottom panel),
as functions of the winding number for $ N=10^3 $ particles and
  $ V_b / \hbar \omega_r = 0 $ (circles), 3 (stars), 
and 5 (triangles). The solid lines correspond to  Eq. (\ref{ev0}) (top) and
 Eq. (\ref{lz00}) (bottom).
}
\label{3le0}
\end{figure}

\begin{figure}
\includegraphics{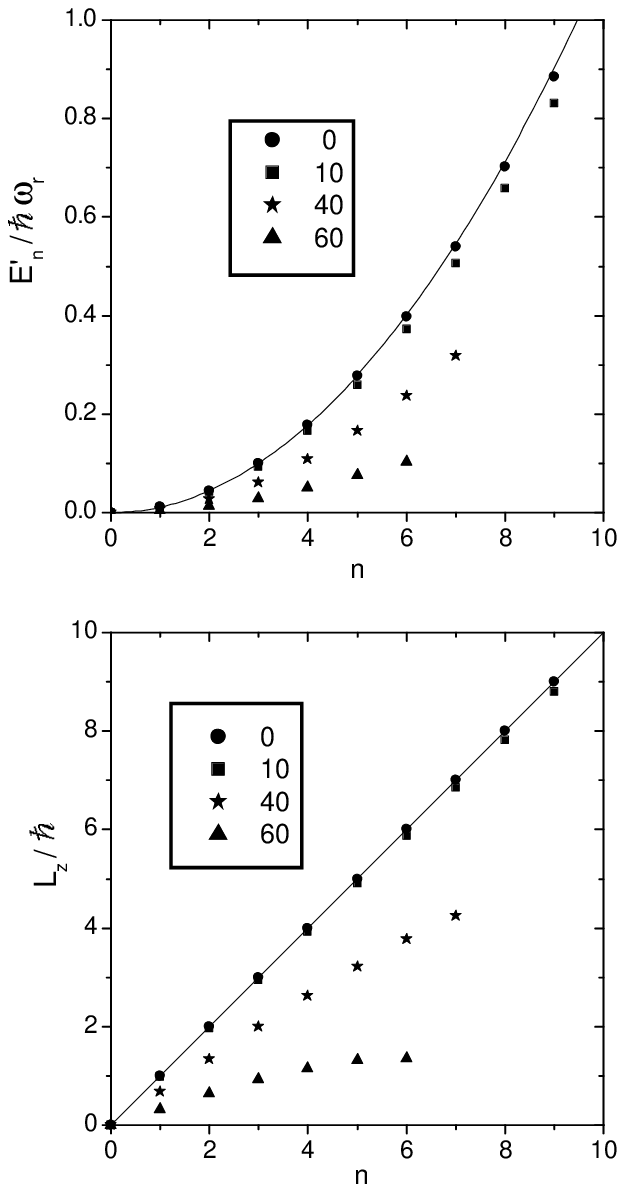}
\caption{  Same as Fig.~\ref{3le0} for the condensate of $ N=10^5 $ particles with
barrier heights
  $ V_b / \hbar \omega_r = 0 $ (circles), 10 (squares), 40 (stars), 
and 60 (triangles). 
}
\label{5le0}
\end{figure}

 In the top panel of Figs.~\ref{3le0} and ~\ref{5le0}
we depict the excitation energy $ E^{\prime}_n= E(n) - E_0$ 
as a function of the winding number. In the case of a
vanishing barrier,
we have also numerically evaluated   the parameter $ K$ of Eq. (\ref{k}),
obtaining  the values 0.01156 and 0.01114
for $ N=10^3 $ and $ N=10^5 $, respectively. Thus, we 
 have drawn the curve arising from formulae (\ref{ev0}),
which shows an excellent agreement 
with the numerically obtained values (circles), as seen from  Figs.~\ref{3le0} and ~\ref{5le0}.
It may also be observed that the energy decreases
for increasing  values of the barrier height,
retaining, however,
the approximately quadratic behavior predicted by Eq. (\ref{ev0}).

\begin{figure}
\includegraphics{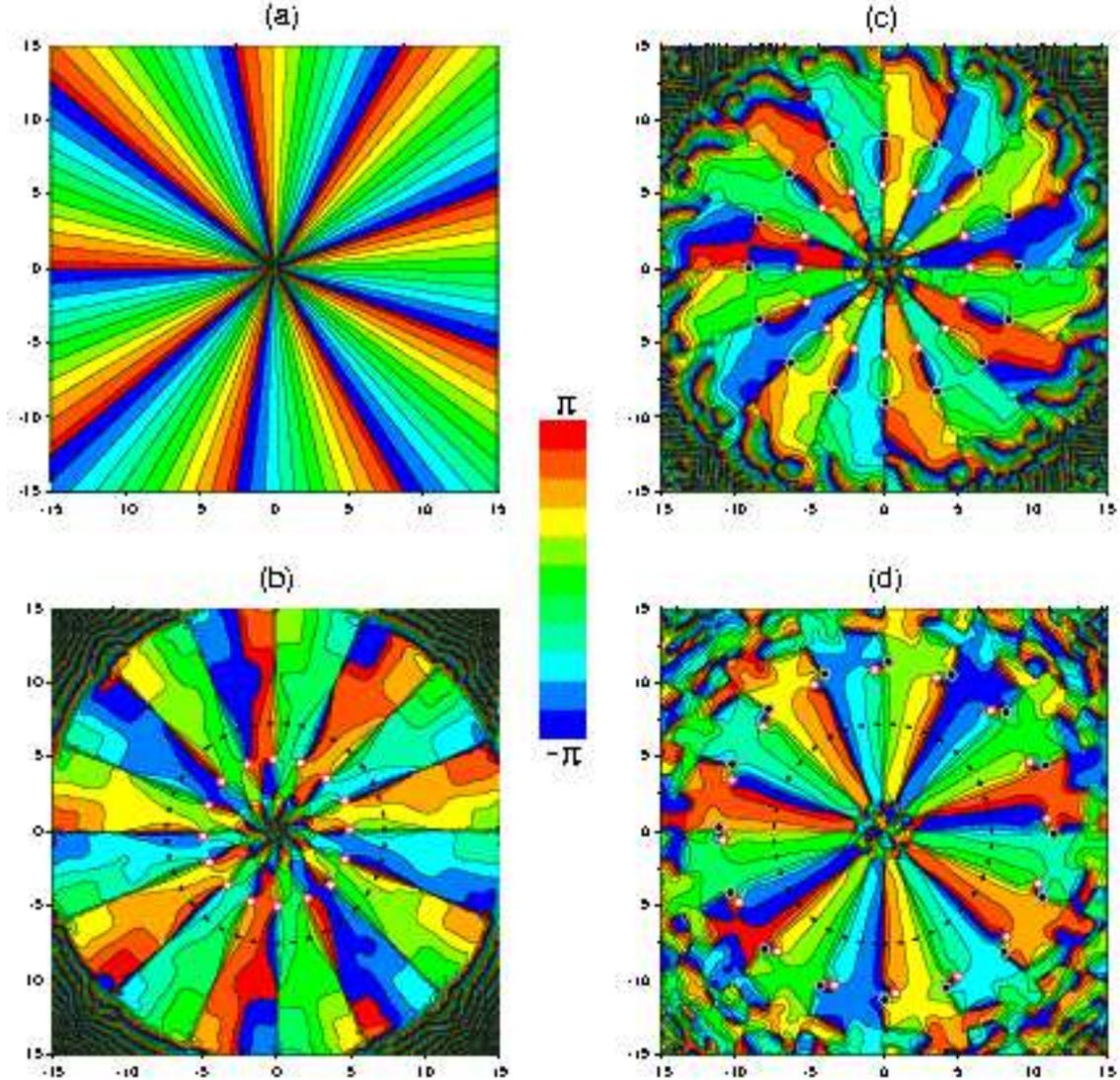}
\caption{ (Color online) Snapshots of the phase distribution in the condensate 
of $10^5 $ particles with a barrier height $ V_b/\hbar \omega_r=40 $ and
an initially imprinted winding number $n=9$. Panels (a) to (d) correspond to
$\omega_{r}t$ = 0, 0.374, 0.796,
and 3.183, respectively. 
White dots in panels (b) to (d) denote vortices
with topological charge $m=+1$, while black dots
in (c) and (d) denote antivortices with $m=-1$. The dashed line circles in (b) and (d)
correspond to circulations yielding initial $n=9$ and final $n'=-7$ winding numbers, respectively. }
\label{32}
\end{figure}

The bottom panel of Figs.~\ref{3le0} and ~\ref{5le0} shows
the angular momentum as a function of the winding number.
Note that for a vanishing barrier, the numerical results correspond to
the straight line predicted by Eq. (\ref{lz00}), while
we may observe that the angular momentum decreases  and curves down
for increasing  values of the barrier height.
In the same sequence, we may also observe a decreasing value of the maximum 
winding number $ \nu $ the system can host.
In fact, we have found that for  a given barrier height,
the time evolution 
 with a moderate dissipation \cite{kasa}
of an initial state with
winding number $ n >\nu $, 
may lead to different processes of vorticity reduction, which involve vortex motion along
the barriers. Such a vortex dynamics may either simply consist of
a vortex escaping phenomenon or, by contrast, it may contain more complex processes such as, 
vortex escaping with the aid of antivortices. 
We describe in the following this last particular process. 
Fig.~\ref{32} shows a series of snapshots in time of the phase distribution
when a winding number $n=9$ is initially imprinted in the 
condensate of 10$^5$ particles and barrier height $ V_b/\hbar \omega_r=40 $ (panel(a)).
We recall that the maximum winding number for this barrier height is $\nu=7$,
as shown in Fig.~\ref{5le0}.
The time evolution first yields a ring of $16$ vortices of topological
charge $m=+1$, denoted as white dots in panel (b), which
move along the barriers from the condensate center outwards. Thus,
the circulation along the dashed line
circle in panel (b) still bears $n=9$, while the same calculation for a circle within the
vortex ring would yield -7. However, to reach the 
condensate border, each vortex should overcome the high density region corresponding to
the potential minimum shown at the left panel of Fig.~\ref{potenr}, which approximately
lies around the dashed line circles shown in Fig.~\ref{32}. 
Such a displacement would produce an 
overall increase in energy that is actually avoided by the generation of an outer ring of antivortices,
 denoted as black dots
in panel (c). This leads to the formation of vortex-antivortex pairs, i.e.,
 dipoles, of lower kinetic energy that are then able to keep on moving towards the
condensate edge, leaving  behind them a pattern of centered vorticity of 
winding number -7. This process is accompanied by a progressive reduction of
the separation of both charges at each dipole. 
In the last panel (d), we may observe that each dipole is about to annihilate itself, while 
the final $n'=-7$ central vorticity pattern becomes evident by following the circulation 
along the dashed line circle. 

Finally, we want to remark that time evolutions with $ n \le \nu $ always lead to 
an identical final winding number
 $ n'= n $,  
with the vorticity
concentrated at the central hole.

\begin{figure}[ht]
\includegraphics{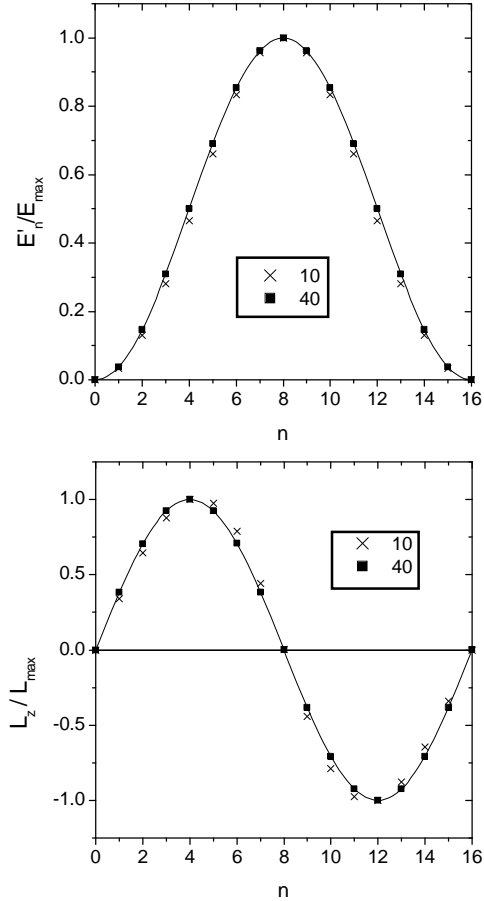}
\caption{ Excitation energy per particle $E'_n$ (top panel)
and angular momentum per particle $L_z=\langle \hat{L}_z\rangle$ (bottom panel),
in units of their maximum values $ E_{max}$ and $ L_{max}$, respectively,
  as functions of the winding
number for  the condensate of $ N=10^3 $ particles,
with barrier heights
 $ V_b/\hbar \omega_r$ = 10 (crosses) and  40  
(squares). The solid lines represent the functions: $[1-\cos(  2 \pi n/ N_c)]/2$ 
(top panel) and $\sin(  2 \pi n/ N_c)$ (bottom panel).
 }
\label{3le1040}
\end{figure}

\begin{figure}[ht]
\includegraphics{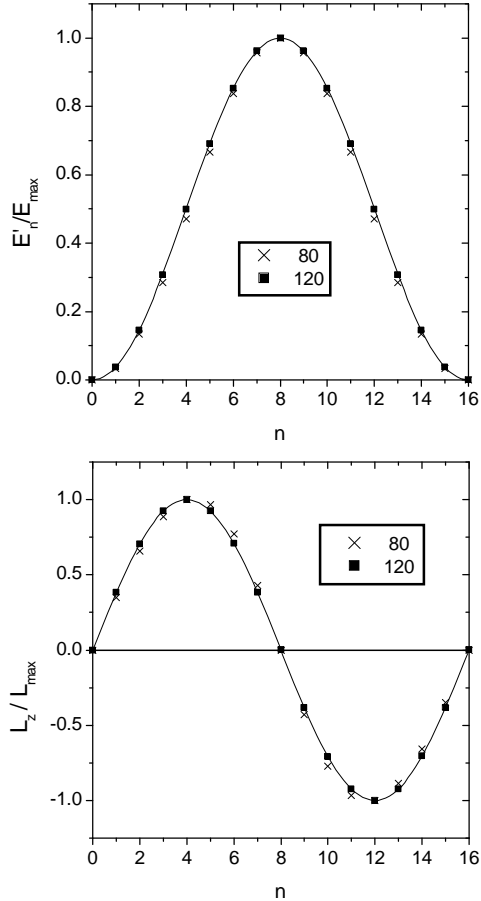}
\caption{Same as Fig.~\ref{3le1040} for the condensate of $N= 10^5$ particles
with barrier heights
 $ V_b/\hbar \omega_r$ = 80 (crosses) and  120  
(squares).}
\label{5le120}
\end{figure}

\subsection{ High barriers}

Now we will study the numerical results obtained for the high-barrier regime which was analytically 
investigated in Sec.~\ref{barrerasaltas}.
We depict in Figs.~\ref{3le1040} and~\ref{5le120}, the excitation energy
 (top panel) and the  angular momentum (bottom panel),
as functions of the winding number 
for the condensates of
$10^3 $ and $ 10^5 $ particles,
respectively.
We may observe that while both magnitudes are
well described by the periodic functions 
obtained in \ref{barrerasaltas}, there still exist for  the lower 
barrier heights, $ V_b/\hbar \omega_r$ = 10 and  80, some
noticeable differences between the numerical values and the predicted curves.

\begin{figure}
\includegraphics{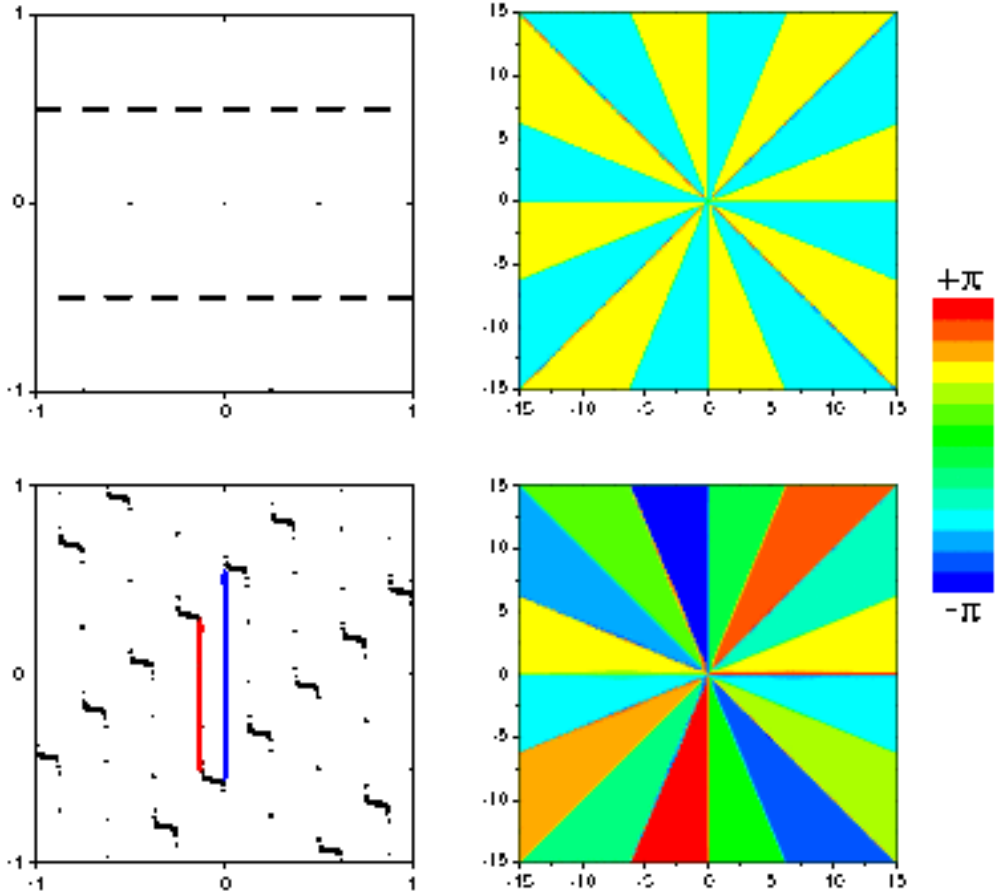}
\caption{(Color online) Phase of the order parameter 
as a function of the angular coordinate (both in units of $\pi$) for $ r_0=6\,l_r$ (left panels),
and phase distribution in the $x$-$y$ plane (right panels),
in the condensate of $10^5$ particles with a barrier height $V_b/\hbar \omega_r=80$.
The imprinted winding number is $n=8$ (top panels) and $n=9$ (bottom panels). The long (blue) vertical arrow
in the left-bottom panel shows a phase difference between neighboring sites larger than $\pi$, which amounts 
to the smaller phase difference denoted by the short (red) arrow. Note also the slightly negative
phase slope within each site, $ \partial \phi /\partial \theta<0$, which yields the final winding number
$n'=9-N_c=-7$ and the corresponding negative angular
momentum, as seen in the bottom panel of Fig.~\ref{5le120}.
}
\label{34}
\end{figure}

\begin{figure}
\includegraphics{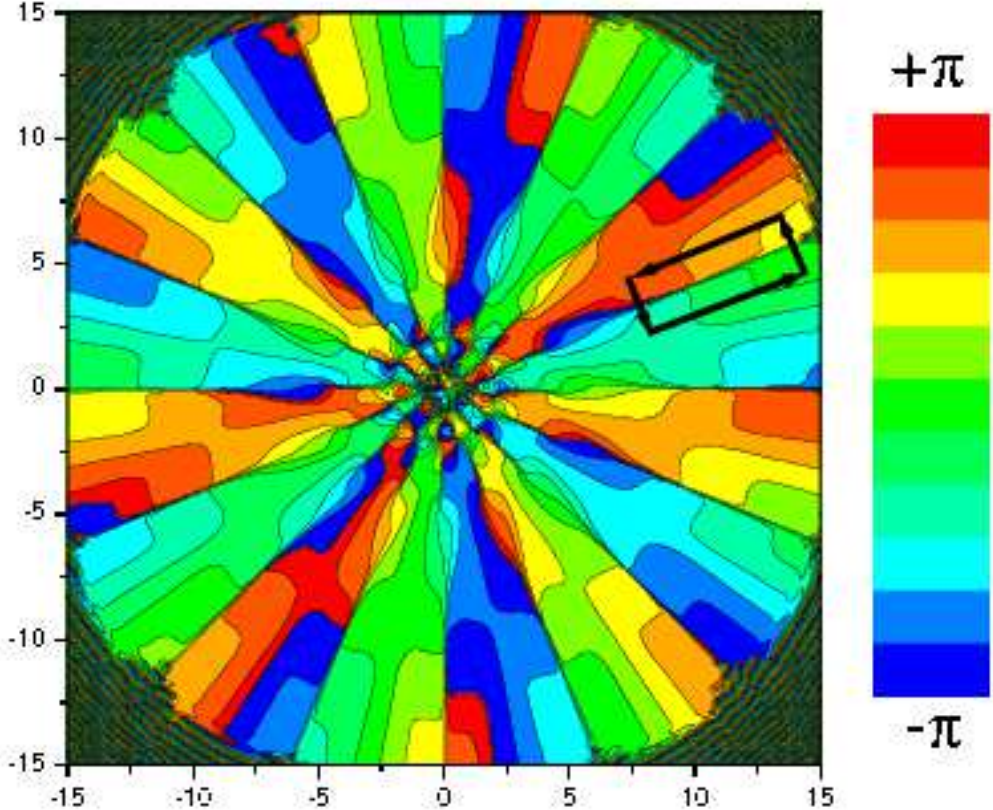}
\caption{(Color online) Snapshot of the phase distribution in 
the condensate of $10^5 $ particles with a barrier height of $ V_b/\hbar \omega_r=80 $
and an initially imprinted winding number $n=9$.
Real-time dissipative evolution at $t= 0.2387\, \omega_{r}^{-1}$ showing
vorticities with a topological charge $m=+1$ spread along each barrier. 
Such a spreading may be appreciated by following the circulation along the
rectangular contour enclosing one of the barriers.  }
\label{33}
\end{figure}

Irrespective of the number of particles, we have observed in this regime
 the  processes
 predicted by P\'erez-Garc\'{\i}a {\it et al.}~\cite{gar07}, 
namely the so-called charge {\it erasure}, which occurs only for an even number of lattice sites, and 
charge {\it inversion}.
In fact, for an imprinted winding number $ n =  N_c/2 = 8$,  
the  phases between  neighboring sites differ in $ \pi$ (top panels of Fig.~\ref{34}), leading to
a vanishing net flux of particles, or in other words, to an erasure of vorticity. 
Notice that such a state 
presents a vanishing angular momentum and a maximum of energy, as shown in Figs.~\ref{3le1040} and~\ref{5le120}.
On the other hand,
charge inversion takes place if $ n > N_c/2 $, as shown in the bottom panels of Fig.~\ref{34}.
In this case, the phase  difference between  neighboring 
sites $\Delta\phi$ becomes larger than $ \pi $, which amounts to an effective phase difference
of $2\pi-\Delta\phi$. Thus, the phase derivative 
$ \partial \phi /\partial \theta$ changes sign, giving rise to
an inverted  current density with a corresponding 
negative angular momentum (see Figs.~\ref{3le1040} and~\ref{5le120}),
and a final  winding number $n' =  n - N_c$.
We have found that repeating the
dissipative real-time simulation of Sec.~\ref{lowb}, but for a higher barrier of 
$ V_b/\hbar \omega_r=80$, 
leads to vortex 
escaping along each of the 16 barriers and
a final state of vorticity $ n'= -7 $, localized within the condensate central hole.
However, it is interesting to notice that in contrast to the low barrier regime 
(see Fig.~\ref{32}), where the vortices
may be viewed as
point charges, 
we observe in Fig.~\ref{33}
a remarkable phenomenon of 
vorticity spreading of the escaping vortices.

Summarizing, we may realize that there exists in this regime, 
irrespective of the number of particles,
$N_c=16$ different stationary states, namely
$ N_c-2 = 14 $  `multivortex'  states corresponding to
final winding numbers $1\leq |n'| \leq N_c/2-1=7$, plus
a single excited state of maximum energy and vanishing particle current corresponding to $ n=N_c/2=8$,
and the ground state ( $n=0$). These results are valid for any even number
of lattice sites. For an odd number of sites, on the other hand,
the only difference would be given by the absence of
the single excited state of maximum energy and vanishing angular momentum,
 i.e., we would have $N_c-1$ 
`multivortex' states corresponding to
final winding numbers $1\leq |n'| \leq (N_c-1)/2$, apart from the ground state with $n=0$ \cite{fer05,gar07}.

\begin{figure}
\includegraphics{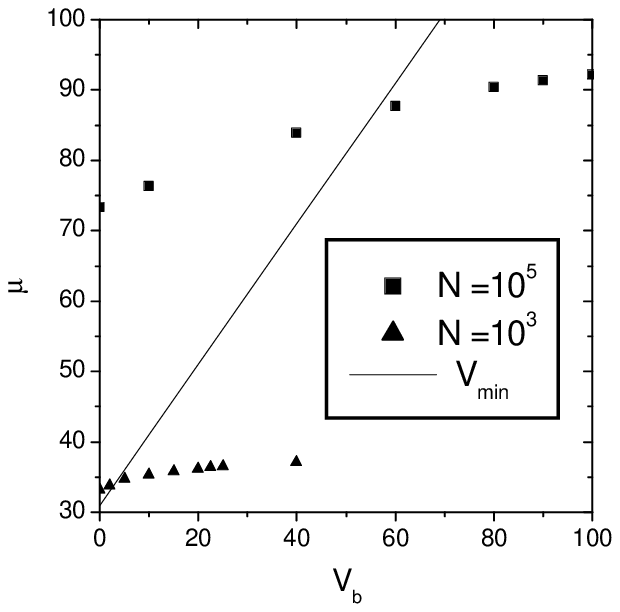}
\caption{ Ground state chemical potential $\mu$ of the condensate containing $10^5$ particles (squares)
and $10^3$ particles (triangles)  as a
function of the barrier height $V_b$. The solid line corresponds to the 
effective potential barrier minima of Fig.~\ref{potenr}.
All quantities are given in units of $\hbar\omega_r$.}
\label{mu5}
\end{figure}

Finally, in Fig.~\ref{mu5} we depict
the chemical potential of the ground state
as a function of the barrier height $V_b$, jointly with
the effective potential barrier minima $ V_{min}$ shown in Fig.~\ref{potenr}.
We notice that $ V_{min}$ approximately
intersects the chemical potentials corresponding to $10^3$ and $10^5$
particles at $ V_b= 5  \, \hbar \omega_r $  and $ V_b= 60 \, \hbar \omega_r $,
respectively. 
Such intersections correspond to transitions between the low- and
high-barrier regimes above described, i.e., they  indicate the barrier height above which
the quantum tunneling regime becomes dominant.

\section{ Summary and concluding remarks}\label{Conclusions} 

We have analyzed
 the angular momentum
and  energy as functions of the winding number
in a ring-shaped lattice, finding two clearly differentiated regimes,
according to the relative value of the barrier height as compared to
the chemical potential. 
On the one hand, we have found that for low barriers,
there exists an almost quadratic dependence on the winding number for the energy and a linear
dependence for the angular momentum, as it is analytically derived in the absence of barriers.
For high barriers, on the other hand, we have found that 
the energy and the angular momentum turn out to be accurately described by sinusoidal functions,
as predicted from a simple analytical model. We have observed
these regimes irrespective of the number of particles,
and a remarkable difference between them consists in that
for the low-barrier regime, metastable vortex states are
obtained up to a maximum winding number $\nu$, which depends on the particle number and the barrier height,
whereas, for high barriers, such states are obtained up to a maximum winding number
which only depends on the number of lattice sites.
 Another difference between both regimes stems from the way in which  vorticity becomes reduced
from an initially imprinted winding number exceeding $ \nu $. 
For a low barrier, this process involves the displacement outwards
along the barriers of positive point charges,
jointly with an eventual generation of negative charges at the condensate
edge. Such antivortices penetrate along the barriers to form dipoles
that,  by diminishing the energy from velocity field, are then able to
cross the maximum density region outwards.
On the other hand, the vorticity reduction in
the large-barrier regime
proceeds through the appearance of stretched positive vorticities along
 each barrier, which eventually escape from the condensate.

To conclude, we believe that the present study 
could also be regarded as a first stage to establish a Bose-Hubbard model for
ring-shaped optical lattices with a large occupation number per site.

\acknowledgments
DMJ acknowledges CONICET for financial support under Grant No. 
PIP 11420090100243.


\begin{thebibliography}{99}

\bibitem{gati} R. Gati and M. K. Oberthaler, J. Phys. B \textbf{40}  R61 (2007).

\bibitem{twowell} M. Asad-uz-Zaman and D. Blume, Phys. Rev. A \textbf{80}, 053622 (2009);
 N. Teichmann, M. Esmann, and C.  Weiss, Phys. Rev. A \textbf{79}, 063620 (2009);
 D. Witthaut,  F. Trimborn,  and S. Wimberger, Phys. Rev. A \textbf{79}, 033621 (2009);
M. T. Martinez,  A.  Posazhennikova,  and J.  Kroha, Phys. Rev. Lett. \textbf{103}, 105302 (2009);
B.  Juli\'a-D\'iaz,  D. Dagnino,  M. Lewenstein,  J. Martorell, and A. Polls, 
 Rev. A \textbf{81},023615 (2010).

\bibitem{zap98}
I. Zapata, F. Sols, and A. J. Leggett,
Phys. Rev. A {\bf 57}, 28 (1998).

\bibitem{zap03}
I. Zapata, F. Sols, and A. J. Leggett,
Phys. Rev. A {\bf 67}, 021603 (2003).

\bibitem{Smerzi1997} A. Smerzi, S. Fantoni, S. Giovanazzi, and S. R. Shenoy, 
Phys. Rev. Lett. \textbf{79}, 4950 (1997).

\bibitem{Raghavan1999} S. Raghavan, A. Smerzi, S. Fantoni, and S. R. Shenoy, 
Phys. Rev. A \textbf{59}, 620 (1999).

\bibitem{Ananikian2006} D. Ananikian and T. Bergeman, Phys. Rev. A \textbf{73}, 013604 (2006).

\bibitem{Albiez2005} M. Albiez, R. Gati, J. F\"olling, S. Hunsmann, M. Cristiani,
 and M. K. Oberthaler, Phys. Rev. Lett. \textbf{95}, 010402 (2005).

\bibitem{scher07} D. R. Scherer, C. N. Weiler, T. W. Neely, and B. P. Anderson,
Phys. Rev. Lett. {\bf 98}, 110402 (2007).

\bibitem{ghosh08} P. Ghosh and F. Sols, Phys. Rev. A {\bf 77}, 033609
 (2008).

\bibitem{gar07}
V. M. P\'erez-Garc\'{\i}a, M. A. Garc\'{\i}a-March,  and A. Ferrando, Phys. Rev. A
 {\bf 75}, 033618 (2007).

\bibitem{fer05}
A. Ferrando, Phys. Rev. E {\bf 72}, 036612 (2005).

\bibitem{ryu07}
C. Ryu, M. F. Andersen, P. Clad\'e, Vasant Natarajan, K. Helmerson, and W. D. Phillips,
 Phys. Rev. Lett. {\bf 99}, 260401 (2007).

\bibitem{wei08}
C. N. Weiler, T. W. Neely, D. R. Scherer, A. S. Bradley, M. J. Davis, and B. P. Anderson,
Nature {\bf 455}, 948 (2008).

\bibitem{cap09}
P. Capuzzi and D. M. Jezek, J. Phys. B {\bf 42}, 145301 (2009).

\bibitem{hend}
K. Henderson, C. Ryu, C. MacCormick, and M. G. Boshier, New J. Phys. {\bf 11}, 043030 (2009).

\bibitem{castin} 
Y. Castin and R. Dum, 
Eur. Phys. J. D {\bf 7}, 399 (1999). 

\bibitem{gros61}
E. P. Gross, Nuovo Cimento {\bf 20}, 454 (1961);
L. P. Pitaevskii, Zh. Eksp. Teor. Fiz.  {\bf 40}, 646 (1961) 
[Sov. Phys. JETP {\bf 13}, 451 (1961)]. 

\bibitem{jez08}
D. M. Jezek, P. Capuzzi, and H. M. Cataldo,
J. Phys. B {\bf 41},  045304 (2008).

\bibitem{cat09}
H. M. Cataldo and D. M. Jezek, Eur. Phys. J. D {\bf 54}, 585 (2009).

\bibitem{ash} N. W. Ashcroft and N. D. Mermin, {\em Solid State Physics},
(Saunders College Publishing, Fort Worth, 1976), chap. 10.

\bibitem{kasa} We have employed a phenomenological
damped GP equation (see, e.g., 
K. Kasamatsu, M. Tsubota, and M. Ueda,
Phys. Rev. A {\bf 67}, 033610 (2003)) with a damping coefficient
$\gamma=0.03$.
%
\end{thebibliography}
\end{document}